\documentclass[iop]{emulateapj}

\usepackage{epsfig}
\usepackage{apjfonts}
\usepackage{amsmath}

\begin{document}

\title{The diffuse gamma-ray  flux associated with sub-PeV/PeV neutrinos from starburst galaxies}
\author{Xiao-Chuan Chang \altaffilmark{1,2},
Xiang-Yu Wang\altaffilmark{1,2}}
\affil{$^1$ School of Astronomy and Space Science, Nanjing University, Nanjing, 210093, China;  xywang@nju.edu.cn \\
$^2$ Key laboratory of Modern Astronomy and Astrophysics (Nanjing University), Ministry of Education, Nanjing 210093, China \\
}
\begin{abstract}
One attractive scenario for the excess of sub-PeV/PeV neutrinos
recently reported by IceCube is that they are produced by cosmic
rays in starburst galaxies colliding with the dense interstellar
medium. {These proton-proton  ($pp$) collisions  also produce
high-energy gamma-rays}, which finally contribute to the diffuse
high-energy gamma-ray background.  We calculate the diffuse
gamma-ray flux with a semi-analytic approach and consider that the
very high energy gamma-rays will be absorbed in the galaxies and
converted into electron-position pairs, which then lose almost all
their energy through synchrotron radiation in the strong magnetic
fields in the starburst region. Since the synchrotron emission
goes into energies below GeV, this synchrotron loss {reduces} the
diffuse high-energy gamma-ray  flux by a factor of about two, thus
leaving more room for other sources to contribute to the gamma-ray
background. For a $E_\nu^{-2}$ neutrino spectrum, we find that the
diffuse gamma-ray flux {contributes} about 20\% of the observed
diffuse gamma-ray background in the 100 GeV range. However, for a
steeper neutrino spectrum, this synchrotron loss effect is less
important, since the energy fraction in absorbed gamma-rays
becomes lower.

\end{abstract}

\section{Introduction}

The IceCube Collaboration reported 37 events ranging from
60\thinspace TeV to 3\thinspace PeV within three years of
operation, correspond to a  $5.7\sigma $ excess over  the
background atmospheric neutrinos and
muons\citep{2014arXiv1405.5303A}.  The observed events can be
fitted by a hard spectra ($E_\nu^{-\Gamma}$) with $\Gamma=2$ and a
cutoff above 2\thinspace PeV, implied by the non-detection of
higher energy events, or alternatively by a slightly softer but
unbroken power-law spectrum with index $\Gamma\simeq
2.2-2.3$\citep{2014arXiv1405.5303A,Anchordoqui2014a}. The best-fit
single-flavor astrophysical flux ($\nu+\bar\nu$) is
$E_\nu^2\Phi_\nu=0.95\pm0.3\times10^{-8}{\rm GeV cm^{-2} s^{-1}
sr^{-1}}$, assuming a $E_\nu^{-2}$ spectrum. The sky distribution
of the these events is consistent with isotropy
\citep{2014arXiv1405.5303A}, implying an extragalactic origin or
possibly Galactic halo origin \citep{2014PhRvD..89j3003T},
although a fraction of them could come from Galactic sources
\citep{Fox2013,2013arXiv1309.4077A,2014PhRvD..89j3002N,Razzaque2013,2013arXiv1311.7188L}.

The source of these IceCube neutrinos is, however, unknown. Jets
and/or cores of active galactic nuclei (AGN), and gamma-ray burst
(GRBs)   have been suggested to be possible sources
\citep{1991PhRvL..66.2697S,2008APh....29....1A,Kalashev2013,Stecker2013,1997PhRvL..78.2292W,He2012,2013PhRvL.111l1102M,Liu&Wang2013},
where the photo-hadronic (e.g., $p\gamma$) reaction is typically
the main neutrino generation process \citep{Winter2013}. On the
other hand, starburst galaxies and galaxy clusters may also
produce PeV neutrinos, mainly via the hadronuclear (e.g., $pp$)
reaction
\citep{2014PhRvD..89h3004L,He2013,2013PhRvD..88l1301M,Tamborra2014,2014arXiv1405.7648A}.
In this paper, we concern the $pp$ process in starburst galaxies,
following earlier works in which remnants of supernovae in general
\citep{2006JCAP...05..003L} or a special class of
supernova--hypernovae \citep{2014PhRvD..89h3004L} in starburst
galaxies accelerate cosmic rays (CRs), which in turn produce
neutrinos by interacting with the dense surrounding medium during
propagation in their host galaxies. {Here we use starburst
galaxies to mean those galaxies that have much higher specific
star formation rates (SSFR) than the "Main Sequence" of
star-forming galaxies.} To produce neutrinos with energy $E_\nu$,
we require a source located at redshift $z$ to accelerate protons
to energies of $50\frac{(1+z)}{2}E_\nu$
\citep{2006PhRvD..74c4018K}. It is suggested that
semi-relativistic hypernova remnants in starburst galaxies, by
virtue of their fast ejecta, are able to accelerate protons to EeV
energies \citep{2007PhRvD..76h3009W,Liu&Wang2012}. Alternatively,
assuming sufficient amplification of the magnetic fields in the
starburst region, normal supernova remnants might also be able to
accelerate protons to $10^{17}$ eV and produce PeV neutrinos
\citep{2013PhRvD..88l1301M}.

In $pp$ collisions, charged and neutral pions are both generated.
Charged pions decay to neutrinos ($\pi ^{+}\rightarrow \nu _{\mu
}\bar{\nu}_{\mu }\nu _{e}e^{+},\,\pi ^{-}\rightarrow
\bar{\nu}_{\mu }\nu _{\mu }\bar{\nu}_{e}e^{-}$), while neutral
pions decay to gamma-rays ($\pi ^{0}\rightarrow \gamma \gamma $).
Thus, PeV neutrinos are accompanied by high-energy gamma-rays. For
a $E_\nu^{-2}$ neutrino spectrum with a single-flavor flux of
$E_\nu^2\Phi_\nu\simeq10^{-8}{\rm GeV cm^{-2} s^{-1} sr^{-1}}$
produced by $\pi^{\pm}$ decay, the  flux of gamma-rays from
$\pi^0$-decay is $2\times10^{-8}{\rm GeV cm^{-2} s^{-1} sr^{-1}}$.
{While low-energy gamma-rays may escape freely from the galaxy,
very high energy (VHE) gamma-rays are absorbed by the soft photons
in starbursts, as starburst galaxies are optically thick to  VHE
gamma rays \citep{2011ApJ...728...11I,2013ApJ...762...29L}.} The
absorbed gamma-rays will convert to electron-positron ($e^{\pm}$)
pairs, which then initiate electromagnetic cascades on soft
background photons, reprocessing the energy of VHE photons to
low-energy emission and contributing to the diffuse gamma-ray
background {\footnote{A simple estimate of the contribution to the
gamma-ray background can be given below if the synchrotron loss of
$e^{\pm}$ pairs formed during the cascades is not taken into
account. Assuming a flat cascade gamma-ray spectrum  (i.e.,
$\Phi_\gamma\propto E_\gamma^{-\alpha_\gamma}$ with
$\alpha_\gamma\sim 2$)
\citep{1975Ap&SS..32..461B,1997ApJ...487L...9C}, the additional
contribution to the gamma-ray background flux by the cascade
emission {can reach as much as} $2\times10^{-8}{\rm GeV cm^{-2}
s^{-1} sr^{-1}}$. Another component that can contribute to the
diffuse gamma-ray emission is $e^{\pm}$ pairs produced
simultaneously with the neutrinos in the $\pi^{\pm}$ decay. The IC
scatterings  of these pairs with infrared photons will produce
very high energy photons, which may be absorbed and finally
contribute to the cascade gamma-ray component with a flux
$\sim10^{-8}{\rm GeV cm^{-2} s^{-1} sr^{-1}}$. Noting that the
observed gamma-ray background {flux observed by Fermi/LAT} is
$\sim10^{-7}{\rm GeV cm^{-2} s^{-1} sr^{-1}}$ at  $\sim100$ GeV
\citep{Abdo2010},  the total diffuse gamma-ray flux associated
with PeV neutrinos in the $pp$ scenario would thus contribute
$\sim50\%$ of the observed diffuse gamma-ray background in the
$\sim100$ GeV range.}}. Detailed calculations show that the
diffuse gamma-ray emission associated with sub-PeV/PeV neutrinos
contribute a fraction $\ga 30\%$ to the gamma-ray background when
the source density follows the star-formation history and
contribute a fraction $\ga40\%$ for non-evolving source density
case \citep{2013PhRvD..88l1301M}. Analytical calculations in
\citet{2014PhRvD..89h3004L}  show that this fraction is about
$\ga40\%$.  Since other sources, such as AGNs
\citep{1996ApJ...464..600S} or truly diffuse gamma-ray emission
from the propagation of ultra-high energy cosmic rays (UHECRs) in
cosmic microwave and infrared background radiation
\citep{1975Ap&SS..32..461B,2009PhRvD..79f3005K,2011PhLB..695...13B,Wang2011},
also contribute significantly to the diffuse gamma-ray background,
a tension between a too high diffuse gamma-ray flux expected from
sources and the observed background flux exists.  {However, the
above calculations do not consider  the synchrotron cooling of
electron-positron ($e^{\pm}$) pairs formed during the cascades. As
we argue in this paper,  this synchrotron loss effect could reduce
the  flux of cascade gamma-ray emission significantly.}

In this paper, we calculate the {accumulated} diffuse gamma-ray
flux by considering the synchrotron loss of  $e^{\pm}$ produced by
the absorbed gamma-rays in the magnetic fields of the starburst
region. For reasonable assumptions of the strength of the magnetic
fields in the starburst region, we find that the synchrotron loss
dominates over the inverse-Compton (IC) scattering loss. Since the
synchrotron emission goes into energies  below GeV, the intensity
of the cascade gamma-ray component is significantly reduced in the
$\sim100$ GeV energy range. In \S 2, we first obtain the
normalization of the  gamma-ray flux of an individual starburst
galaxy with the observed neutrino flux. In \S 3, we calculate the
gamma-ray flux after considering the synchrotron loss. Then in \S
4, we calculate the {accumulated} gamma-ray flux from the
population of starburst galaxies. Finally we give our conclusions
and discussions in \S 5.

\section{Normalization} To calculate the {accumulated} diffuse gamma-ray
flux, we need to know the the gamma-ray flux from individual
starburst galaxies and then integrate over contributions by all
the starburst galaxies in the Universe. The  gamma-ray flux of an
individual galaxy can be then calibrated with the observed
neutrino flux, assuming that all the observed neutrinos by IceCube
are produced by starburst galaxies.

As the neutrino flux is linearly proportionally to the cosmic-ray
proton flux and the  efficiency  in transferring the proton energy
to secondary pions,  the single-flavor ($\nu+\bar\nu$) neutrino
luminosity per unit energy from a starburst galaxy is expressed by
\begin{equation}
L _{\nu _{i}}=\frac{dN_{\nu _{i}}}{dtdE_\nu}\propto f_{\pi
}L_{p}E_{\nu}^{-p},
\end{equation}
where $i=(e,\mu,\tau)$,  $L_{p}$ is the CR proton flux at some
fixed energy and $f_\pi$ is  the pion production efficiency. $p$
is the index of the proton spectrum ($dn/dE_p\propto E_p^{-p}$),
and we assume $p=2$, as expected from Fermi acceleration, unless
otherwise specified. Starbursts have very high star formation
rates ($SFR$) of massive stars. Massive stars produce supernova or
hypernovae, which accelerate CRs in the remnant blast waves, so we
expect that $L_{p}\propto SFR$. On the other hand, the radiation
of a large population of hot young stars are absorbed by dense
dust and reradiated in the infrared band, so $L_{\rm TIR}\propto
SFR$, where $L_{\rm TIR}$ is the total infrared luminosity of the
starburst galaxy. Thus we expect that $L_p\propto L_{\rm TIR}$, so
\begin{equation}
L _{\nu _{i}}(E_{\nu}, L_{\rm TIR})=C_1f_{\pi }\frac{L_{\rm
TIR}}{L_\odot}\left(\frac{E_{\nu}}{1{\rm GeV}}\right)^{-p},
\end{equation}
where $C_1$ is the normalization factor and $L_\odot$ is the
bolometric luminosity of the Sun. The {pionic} efficiency is
$f_{\pi }=1-\exp (-t _{esc}/t _{loss})$, where $t_{loss}$ is the
energy-loss time for $pp$ collisions and $t _{esc}$ is the escape
time of protons. $t _{loss}=(0.5\,n\,\sigma _{pp}\,c)^{-1}$, where
the factor 0.5 is inelasticity, $n$ is the particle density of gas
and $\sigma _{pp}$ is the inelastic proton collision cross
section, which is insensitive to proton energy. Introducing a
parameter $\Sigma_g=m_pnR$ as the surface mass density of the gas
(where $R$ is the length scale  of the starburst  region), the
energy loss time is
\begin{equation}
t_{loss}=7\times10^{4}{\rm
yr}\frac{R}{500pc}\left(\frac{\Sigma_g}{1{\rm g
cm^{-2}}}\right)^{-1}.
\end{equation}
We use $R=500{\rm pc}$ as the reference value \footnote{For
comparison, we also consider $R=200{\rm pc}$ and $R=1{\rm kpc}$ in the following calculations.} because neutrinos are dominantly produced by high
redshift starbursts, as implied by the star-formation history, and
high redshift starburst typically have $R\sim 500{\rm pc}$
\citep{2006ApJ...640..228T}. There are two ways for CRs to escape
from a galaxy. One is the advective escape via a galactic wind and
the other is the diffusive escape. The advective escape time is
\begin{equation}
t_{adv}=R/v_w=4.5\times10^5 {\rm yr}
\frac{R}{500pc}\left(\frac{v_w}{1000{\rm km s^{-1}}}\right)^{-1},
\end{equation} where  $v_w$ is the velocity of galactic wind. One can see that
protons lose almost all their energy in dense starburst galaxies
with $\Sigma_g\ga 0.15(v_w/1000{\rm km s^{-1}}){\rm g cm^{-2}}$,
if the advective time is shorter than the diffusive escape time.
Little is known about the diffusive escape time in starbursts. The
diffusive escape time can be estimated as $t_{diff}=R^2/2D$, where
$D=D_0(E/E_0)^\delta$ is  the diffusion coefficient, $D_0$ and
$E_0$ are normalization factors, and $\delta=0-1$ depending on the
spectrum of interstellar magnetic turbulence. {As in
\citet{2014PhRvD..89h3004L} and \citet{2013PhRvD..88l1301M}, we
assume a lower diffusion coefficient $D_0\la10^{27} {\rm cm^2
s^{-1}}$ at $E_0=3{\rm GeV}$ for  starburst galaxies, because the
magnetic fields in nearby starburst galaxies such as M82 and
NGC253 are observed to be  $100$ times stronger than in our Galaxy
and the diffusion coefficient is expected to scale with the CR
Larmor radius in the case of Bohm diffusion. The fact that no
break in the GeV-TeV gamma-ray spectrum up to several TeV from
HESS observations also suggests a small diffusion coefficient in
starburst galaxies \citep{2012ApJ...757..158A}.}  The diffusive
escape time is thus
\begin{equation}
t_{diff}=10^5{\rm yr}(\frac{R}{500pc})^2(\frac{D_0}{10^{27} {\rm
cm^2 s^{-1}}})^{-1}(\frac{E_p}{60{\rm PeV}})^{-0.3}
\end{equation}
for $\delta=0.3$ \citep{2011ApJ...729..106T}. Alternatively,
\citet{2011ApJ...734..107L} assume that CRs stream out the
starbursts at the average Alfv\'{e}n speed $v_A=B/\sqrt{4\pi m_p
n}$, so
\begin{equation}
t_{diff}=2.2{\rm
Myr}\left(\frac{R}{500pc}\right)^{1/2}\left(\frac{\Sigma_g}{1{\rm
g cm^{-2}}}\right)^{1/2} \left(\frac{B}{2{\rm mG}}\right)^{-1},
\end{equation}
where $B$ is the magnetic field in the starburst region. The
escape time is the minimum of the advective time and the diffusive
time, i.e. $t_{esc}=\min(t_{adv}, t_{diff})$.

The {accumulated} neutrino  flux can be estimated by summing up
the contribution by all starburst galaxies throughout the whole
universe, i.e.
\begin{equation}
E_\nu^{2}\Phi_{\nu_i}^{\rm accu} =\frac{E_\nu^{2}c}{4\pi }
\int_{0}^{z_{\max }}\int_{L_{TIR,\min }}^{L_{TIR,\max }}\frac{\phi
(L_{\rm TIR},z)L_{\nu _{i}}[(1+z)E_\nu,L_{\rm
TIR}]}{H_{0}\sqrt{(1+z)^{3}\Omega _{M}+\Omega _{\Lambda }}}
dL_{\rm TIR}dz,
\end{equation}
where $\phi (L_{\rm TIR},z)$ represents the luminosity function of
starburst, $H_{0}=71{\rm \,km\,s^{-1}\,Mpc^{-1}}$, $\Omega
_{M}=0.27$, $\Omega _{\Lambda }=0.73$.  Herschel PEP/HerMES has
recently provided an estimation of the IR-galaxy luminosity
function for separate classes \citep{2013MNRAS.432...23G}, which is a modified-Schechter
function
\begin{equation}
\phi (L_{\rm TIR})=\phi ^{\ast }\left( \frac{L_{\rm TIR}}{L^{\ast
}}\right) ^{1-\alpha
}\exp \left[-\frac{1}{2\sigma ^{2}}\log _{10}^{2}\left( 1+\frac{L_{\rm TIR}}{L^{\ast }}%
\right) \right]\,.
\end{equation}
For starburst galaxies, $\alpha =1.00\pm 0.20$, $\sigma =0.35\pm
0.10$, $\log _{10}(L^{\ast }/L_{\odot})=11.17\pm 0.16$, {$\log_{
10}(\phi ^{\ast }/{\rm Mpc^{-3}dex^{-1}})=-4.46\pm 0.06$}.
$L^{\ast }$ evolves with redshift as $\propto (1+z)^{k_{L}}$ with
$k_{L}=1.96\pm 0.13$. For $\phi^{\ast }$, it evolves as $\propto
(1+z)^{k_{\rho ,\,1}}$ when $z<z_{b,\,\rho }$, and as $\propto
(1+z)^{k_{\rho ,\,2}}$ when $z>z_{b,\,\rho }$, with $k_{\rho
,\,1}=3.79\pm 0.21$, $k_{\rho ,\,2}=-1.06\pm 0.05$ and
$z_{b,\,\rho }=1$. We set $z_{\max }=4$  and the luminosity ranges
from $10^{10}\,L_{\odot}$ to $10^{14}\,L_{\odot}$.

We assume all the observed neutrinos are produced by starburst
galaxies, so we can normalize the {accumulated} diffuse neutrino
flux in Eq. (7) with the observed flux at PeV energy, i.e.
\begin{equation}
E_\nu^{2}\Phi_{\nu_i}^{\rm accu}|_{E_\nu=1\,{\rm PeV}} =
0.95\times 10^{-8}\,{\rm GeV\,cm^{-2}\,s^{-1}\,sr^{-1}}.
\end{equation}
Using the central values of the above parameters in the luminosity
function (e.g. $\alpha =1.00$, $\sigma =0.35$ and $\log
_{10}(L^{\ast }/L_{\odot})=11.17$), we get $C_{1}=4.1\times
10^{21}{\rm eV^{-1}s^{-1}}$ for $p=2$. Varying the values of these
parameters within their error, the value of $C_1$  changes within
a factor of a few. However, we  note that the result of the
{accumulated} gamma-ray flux, as calculated below in \S 4, remains
almost unchanged.  We also find that the spectrum of the
{accumulated} neutrino flux, as given by Eq.(7), is flat ($\propto
E_\nu^{-2}$) for either choice of the diffusive escape times in
Eq.(5) or Eq.(6).

For $pp$ collisions, the energy flux in $\pi^0$-decay gamma-rays
is  $E_{\gamma }^{2}L_{\gamma }\propto(1/3)E_{p}Q_{p}$, where
$Q_{p}$ represents the differential CR energy flux. Since the
neutrino flux is $E_{\nu _{i}}^{2}L_{\nu}\propto (1/6)E_{p}Q_{p}$,
the gamma-ray luminosity per unit energy resulted from the
$\pi^0$-decay is related to the neutrino luminosity by
\begin{equation}
E_{\gamma }^{2}L_{\gamma }\approx 2E_{\nu}^{2}L_{\nu
_{i}}|_{E_{\nu}=0.5E_{\gamma }}.
\end{equation}
Thus the $\pi^0$-decay photon luminosity per unit energy of a
single starburst should be
\begin{equation}
L_{\gamma }=8.2\times 10^{21}f_\pi \frac{L_{\rm
TIR}}{L_\odot}(\frac{E_\gamma}{1{\rm GeV}})^{-2} {\rm eV^{-1}
s^{-1}}
\end{equation}
for $p=2$.

\section{Gamma-ray flux from one single starburst galaxy}
The $pp$ process   produces neutral pions $\pi^0$ and charged
pions $\pi^{\pm}$. The neutral pion decay produces gamma-rays.
Besides, $e^{\pm}$ pairs from  the decay of $\pi^{\pm}$, can also
produce gamma-rays through IC scattering off the soft photons in
the starburst region, mainly the infrared photons (in competing
with the synchrotron radiation in the magnetic field). While the
low-energy gamma-rays can escape and contribute to the diffuse
gamma-ray background, the VHE gamma-rays will be absorbed due to
the dense photon field in the starburst region. The absorbed
gamma-rays will be converted into $e^{\pm}$ pairs, which initiate
cascades through interactions with the soft photons. As the VHE
gamma-rays cascade down to enough low energies, they can escape
out of the starburst and also contribute to the diffuse gamma-ray
background. Below we consider the separate contributions by the
absorbed gamma-rays and the unabsorbed ones.

\subsection{Absorption of high-energy gamma-rays in starbursts}
Rapid star formation process makes an intense infrared photon
field in the starburst. The VHE gamma-ray photons may be absorbed
by background photons and generate $e^{\pm}$ pairs. The optical
depth is $\tau _{\gamma \gamma }(E_{\gamma })=\int\sigma _{\gamma
\gamma }(\varepsilon ,E_{\gamma }) \,n(\varepsilon
)\,R\,d\varepsilon $, where $\sigma _{\gamma \gamma }$ is the
cross section for pair-production
\citep{1971MNRAS.152...21B,2012ApJ...758..101S} and $n(\varepsilon
)$ is the number density the infrared background photons. We adopt
the optical-infrared spectral energy models for the nuclei of
starburst galaxies developed by \citet{2007A&A...461..445S} and
take Arp 220 as a  template for starbursts to calculate the
absorption optical depth. Using Arp220 as a template is because
its infrared luminosity is close to luminosities of those
starburst galaxies that contribute dominantly to the diffuse
neutrino background (see \S 4). {We find that the absorption
energy does not change too much if we use M82 as a template
instead.} For simplicity, we assume all starbursts have the same
photon spectra as Arp 220. With such simplifications, the optical
depth $\tau _{\gamma \gamma }(E_{\gamma })$ in a starbusrt depends
only on its total infrared luminosity $L_{\rm TIR}$. {Fig. 1 shows
the corresponding absorption energy $E_{cut}$ (where $\tau
_{\gamma \gamma }=1$) as a function of the total infrared
luminosity of starburst galaxies for different choices of
$R=200{\rm pc}$,$R=500{\rm pc}$ and $R=1{\rm kpc}$.} It can be
seen that photons with energies above $1-10{\rm TeV}$ will be
absorbed inside the starburst region.

\subsection{Absorbed gamma-rays} The $\pi^0$ decay gamma-rays
above $E_{cut}$ will be absorbed in the soft photons in the
starbursts and produce pairs, each carrying  about half energy of
the initial photons ($\gamma \gamma \rightarrow e^{+}e^{-}$).
These pairs, together with the pairs resulted from $\pi^{\pm}$
decay, will cool through synchrotron and IC emission. The
synchrotron photon energy is $E_{syn}=50 (E_{e}/1{\rm PeV})^2
(B/1{\rm mG}){\rm MeV}$, where $E_e$ is the energy of $e^{\pm}$
pairs and $B$ is the magnetic field in the starburst region. Under
reasonable assumptions about the magnetic field in the starburst
region, the synchrotron photon energy is below GeV energies, so
their contribution to the diffuse gamma-ray flux at $\sim100$ GeV
is negligible. Note that the most constraining point of the
gamma-ray background is around 100 GeV, since the observed
gamma-ray background flux decreases with the energy. The IC photon
energy is $E^{IC}_{\gamma}\sim \min [\gamma_e
^{2}\varepsilon_{b},E_e]$, where  $\gamma_e$ is the Lorentz factor
of the pair and $\varepsilon_{b}$ is the energy of the soft
background photons. Approximating the soft photons as infrared
photons with energy $\varepsilon_{b}=0.01{\rm eV}$, The IC photons
have energy of $E^{IC}_{\gamma}=5{\rm TeV}(E_e/10{\rm
TeV})^2(\varepsilon_{b}/0.01{\rm eV})$. The IC photons above
$E_{cut}$ will be  absorbed and an electro-magnetic cascade will
be developed. This cascade process transfers the  energy of
absorbed gamma-rays to lower and lower energies, until the
secondary photons can escape from absorption. The escaped
gamma-ray photons can then contribute to the diffuse gamma-ray
background.  In order to calculate the fraction of the energy loss
of $e^{\pm}$ pairs transferred to the cascade emission, we need to
know the energy loss fraction of $e^{\pm}$ through the IC
emission.

\subsubsection{Synchrotron vs IC loss}
Energy loss timescales are crucial for assessing the fraction of
energy loss in these two processes. \citet{2006ApJ...645..186T}
argued that in order for starburst galaxies to fall on the
observed FIR-radio correlation, the synchrotron cooling time in
starbursts must be shorter than the IC cooling time and the escape
time for relativistic electrons. The reason is that, if this
constraint is not satisfied, any variation in the ratio between
the magnetic field and photon energy density ($U_B/U_{ph}$) would
lead to large changes in the fraction of cosmic ray electron
energy radiated via synchrotron radiation. A linear FIR-radio
correlation would then require significant fine tuning. The
synchrotron timescale depends on the magnetic field in the
starburst region. The magnetic fields in starbursts are not
well-understood and we parameterize the strength of the
magnetic field in terms of the gas surface density $\Sigma_g$ with the following three  scalings : \\
i) First, in the assumption that the magnetic energy density
equilibrates with the total hydrostatic pressure of the
interstellar medium, $B\simeq (8\pi^2 {\rm G})^{1/2}\Sigma_g$
\citep{2006ApJ...645..186T}, where ${\rm G}$ is the gravitational
constant. Fields strength as large as this equipartition are
possible if the magnetic energy density equilibrates with the
turbulent energy density of the ISM. The measurements of Zeeman
splitting associated with OH megamasers for eight galaxies suggest
that the magnetic energy density in the interstellar medium of
starburst galaxies is indeed comparable
to their hydrostatic gas pressure \citep{2014ApJ...780..182M}.\\
ii)$B\propto \Sigma_g^{0.7}$ is sometimes assumed, motivated by
setting the magnetic energy density equal to the pressure in the
ISM produced by star formation ($P_{\rm SF}$). Because of $P_{\rm
SF}\propto \Sigma_{\rm SFR}$ (where $\Sigma_{\rm SFR}$ is the star
formation rate per unit area) and  the Schmidt scaling law for
star-formation $\Sigma_{\rm SFR}\propto \Sigma_g^{1.4}$
\citep{1998ApJ...498..541K}, the scaling $B\propto \Sigma_g^{0.7}$ follows.\\
iii)The third is the minimum energy magnetic field case $B\propto
\Sigma_g^{0.4}$, which is obtained using the observed radio flux
and assuming comparable cosmic-ray and magnetic energy densities
\citep{2006ApJ...645..186T}. Although such a low magnetic field is
not favored, as argued in various aspects
\citep{2006ApJ...645..186T,2014ApJ...780..182M}, we keep this case
in the calculation just to illustrate the difference when a low
magnetic field strength is considerred.
 \\
Thus, the strength of the magnetic fields can be summarized as
\citep{2006ApJ...645..186T,2010ApJ...717..196L}
\begin{equation}
\frac{B}{\mu G}=\left\{
\begin{array}{cc}
2000\left( \frac{\Sigma _{g}}{g\,cm^{-2}}\right) & (B\propto \Sigma _{g}) \\
400\left( \frac{\Sigma _{g}}{g\,cm^{-2}}\right) ^{0.7} & (B\propto
\Sigma
_{g}^{0.7}) \\
150\left( \frac{\Sigma _{g}}{g\,cm^{-2}}\right) ^{0.4} & (B\propto
\Sigma
_{g}^{0.4})%
\end{array}%
\right.
\end{equation}%
\newline
As the  gas surface density $\Sigma _{g}$ scales with the surface
density of star formation rate as $\Sigma _{g}\propto\Sigma
_{SFR}^{0.71}$ and the star formation rate ($\pi R^2 \Sigma
_{SFR}$)  scales linearly with the total infrared luminosity
$L_{TIR}$, we get the relation (assuming a constant star-formation
radius $R$)
\begin{equation}
\frac{\Sigma _{g}}{\rm g\,cm^{-2}}\simeq 3.6\left(
\frac{L_{TIR}}{10^{12}L_\odot}\right) ^{0.71}.
\end{equation}

The synchrotron energy loss timescale of $e^{\pm }$ is $t_{syn}=
6\pi m_ec^{2}/(c\sigma _{T}B^{2}\gamma_e)$
\citep{1979rpa..book.....R},
while the IC energy loss timescale of relativistic $e^{\pm }$, for
a graybody approximation for the soft background photons, is given
by \citep{2010NJPh...12c3044S}
\begin{equation}
t_{IC}\approx \frac{3m_{e}c^{2}}{4c\sigma _{T}U_{ph}}\frac{\gamma
_{K}^{2}+\gamma_e ^{2}}{\gamma_e \gamma _{K}^{2}}
\end{equation}
where $U_{ph}\approx L/(2\pi R^{2}c)$  is the energy density of
the soft photons, $\gamma _{K}\approx 4.0\times
10^{7}(T/40K)^{-1}$, $T$ is the temperature of the graybody
radiation field  and $\gamma_e$ is the Lorentz factor of  $e^{\pm
}$. Note that Eq.(14)  applies to both the Thomson scatterings and
the  scatterings in the Klein-Nishina regime when $\gamma_e$ is
very high.

The  fraction of the energy loss  through synchrotron radiation is
$t_{syn}^{-1}/\left( t_{syn}^{-1}+t_{IC}^{-1}\right) $ for one
electron or positron of a particular energy. Since this fraction
is a function of the energy of $e^{\pm}$, we integrate it over a
proper energy range to estimate the  fraction of the total energy
loss of $e^{\pm}$ through synchrotron radiation  in the energy
range, which is
\begin{equation}
r\approx \frac{\int_{E_{cut}/2}^{E_{\max }/2}\frac{t_{syn}^{-1}}{%
t_{syn}^{-1}+t_{IC}^{-1}}E_e^{-p+1}dE_e}{\int_{E_{cut}/2}^{E_{\max
}/2}E_e^{-p+1}dE_e },
\end{equation}
where $E_{max}=4{\rm PeV}$ is used for the maximum energy of
$e^{\pm}$, corresponding to a maximum neutrino energy of $2{\rm
PeV}$. We show, in Fig.2, this fraction for starburst galaxies
with different $L_{TIR}$. The black, red and blue lines represent,
respectively, the $B\propto \Sigma _{g}$, $B\propto \Sigma
_{g}^{0.7}$ and $B\propto \Sigma _{g}^{0.4}$ cases. It shows that
for both the $B\propto \Sigma _{g}$ and $B\propto \Sigma
_{g}^{0.7}$ cases, the synchrotron loss constitute a fraction
$\ga90\%$ of the total energy loss. There are two factors that
leads to such a large synchrotron loss fraction: 1) the magnetic
energy density is larger than the photon energy density in these
cases; 2) the Klein-Nishina effect for $e^{\pm}$ above $\sim
20(T/40{\rm K})^{-1}{\rm TeV}$ \citep{2013ApJ...762...29L} causes
the IC energy loss time to increase with $\gamma_e$, while the
synchrotron loss time continues to fall as $\gamma_e^{-1}$.

\subsubsection{Cascade gamma-rays}
While the synchrotron loss energy goes into low-energy emission
and thus does not contribute to the diffuse high-energy gamma-ray
background, the IC loss energy  will cascade down to the relevant
energy range of the diffuse gamma-ray emission. If the cascade
develops sufficiently, the spectrum of the cascade emission has a
nearly universal form of
\begin{equation}
L_{cas}\propto \left\{
\begin{array}{cc}
E_\gamma^{-1.5} & (E_\gamma<E_{\gamma,b}) \\
E_\gamma^{-\alpha_\gamma} & (E_{\gamma,b}<E_\gamma<E_{cut})
\end{array}
\right.
\end{equation}
where $E_{cut}$ is  the absorption cutoff energy,
$E_{\gamma,b}\approx
(4/3)(E_{cut}/2m_{e}c^{2})^{2}\varepsilon_{b}$ is the break energy
corresponding to $E_{\gamma,cut}$, and $\alpha_{\gamma}\simeq 2$
typically \citep{1975Ap&SS..32..461B,1997ApJ...487L...9C}. The
spectrum above $E_{cut}$ decreases rapidly as $e^{-\tau _{\gamma
\gamma }}$. The normalization of the cascade emission spectrum is
determined by equating the total cascade energy with the total IC
energy loss above $E_{cut}$, i.e.
\begin{equation}
\int E_\gamma L_{cas}dE_\gamma\simeq\frac{3}{2}(1-r)E_{abs},
\end{equation}
where $E_{abs}=\int_{E_{cut}/2}^{E_{\max }/2}E_\gamma L_\gamma
dE_\gamma$ is the energy of all $\pi^0$-decay gamma-rays above
$E_{cut}$ and the factor $\frac{3}{2}$ represents the sum
contributions by the $\pi^0$-decay gamma-rays and the gamma-rays
from IC scattering of $\pi^\pm$-decay $e^{\pm }$, the latter of
which has a  flux about half  of that of $\pi^0$-decay gamma-rays.

\subsection{Unabsorbed gamma-rays} The gamma-rays from $\pi^0$
decay with energies below $E_{cut}$ will escape out of the
starburst galaxies and contribute directly to the diffuse
gamma-ray background. This flux is model-independent and readily
obtained from the observed neutrino flux by using Eq.(10). For a
flat neutrino spectrum, the differential flux of $\pi^0$-decay
gamma-rays below $E_{cut}$ is $E_\gamma^2
\Phi_{\gamma}\simeq2\times 10^{-8}{\rm GeV cm^{-2} s^{-1}
sr^{-1}}$.

As we pointed out before,  the electrons and positrons
accompanying with the neutrino production in the $\pi^{\pm}$ decay
also contribute to the diffuse gamma-rays through IC scattering of
soft background photons. These $e^{\pm}$ pairs cool by both
synchrotron and IC emission.  We denote the flux contributed by IC
process with $L_{\pi^{\pm},IC }$, which is important only when the
magnetic energy density in the starburst region is low. Therefore
the flux of the unabsorbed gamma-rays is
\begin{equation}
L_{\gamma,un}=(L _{\gamma }+L _{\pi^{\pm},IC })e^{-\tau _{\gamma
\gamma }}
\end{equation}

Adding the luminosity of the unabsorbed gamma-rays and that of the
cascade gamma-rays, we get the total gamma-ray photon luminosity
emitted from one starburst galaxy
\begin{equation}
L _{total}=L_{\gamma,un}+L _{cas}.
\end{equation}

\section{The {accumulated} diffuse gamma-ray flux}
Once the gamma-rays escape out of the starburst galaxy, they  will
further interact with the extragalactic infrared and microwave
background photons, and similar cascades are formed for VHE
gamma-rays. In this case, however, $E_{cut}$ is different and
dependent on the redshift distribution of  starburst galaxies. The
cascade emission spectrum has the same form as equation (16).
Taking such cosmic  cascades into consideration, the accumulated
gamma-ray flux includes two parts, one is from the direct source
contribution and the other is from the intergalactic cascades,
i.e.
\begin{equation}
\Phi_{\gamma}^{\rm accu}=\Phi_{\gamma}^{\rm
sour}+\Phi_{\gamma}^{\rm cas},
\end{equation}
where $\Phi_{\gamma}^{\rm sour}$ represents the direct source
contribution by starburst galaxies and $\Phi_{\gamma}^{\rm cas}$
represents the flux from the intergalactic cascades.

The direct source contribution can be obtained by integrating the
contributions of all starburst galaxies in the universe over their
redshift and luminosity range, i.e.,
\begin{equation}
E_\gamma^{2}\Phi_{\gamma}^{\rm sour} =\frac{E_\gamma^{2}c}{4\pi }
\int_{0}^{z_{\max }}\int_{L_{TIR,\min }}^{L_{TIR,\max }}\frac{\phi
(L_{\rm TIR},z)L' _{total}[{(1+z)E_\gamma
}]}{H_{0}\sqrt{(1+z)^{3}\Omega _{M}+\Omega _{\Lambda }}} dL_{\rm
TIR}dz ,
\end{equation}
where $L'_{total}[{(1+z)E_\gamma  }] = {L_{total}}[(1 +
z){E_\gamma }]{e^{-\tau' _{\gamma\gamma }({E_\gamma },z)}}$ and
$\tau'_{\gamma\gamma}$ is the absorption optical depth due to the
extragalactic infrared and microwave background photons.

 The {accumulated} diffuse gamma-ray flux are shown in
Figures 3-5 for the cases of $R=500{\rm pc}$, $R=200{\rm pc}$ and
$R=1{\rm kpc}$ respectively. It can be seen that, the diffuse
gamma-ray flux in the case  considering the synchrotron loss  is
less than half of that in the case neglecting the synchrotron
loss. This is because that the cascade component resulted from the
absorbed gamma-rays is strongly suppressed due to the synchrotron
radiation of the $e^{\pm}$. The {accumulated} diffuse gamma-ray
flux after considering the synchrotron loss effect contributes
$\sim20\%$ of the observed diffuse gamma-ray background by
Fermi/LAT at $\sim 100 {\rm GeV}$ for both $B\propto \Sigma_g$ and
$B\propto \Sigma_g^{0.7}$ cases.

For a steeper neutrino spectrum, since the energy fraction in
absorbed gamma-rays above $E_{cut}$ is lower, this effect is
expected to be less important. As in \citet{2013PhRvD..88l1301M},
we study the allowed range of the spectral index by the observed
gamma-ray background data. We find that the  index must be
$p\la2.18$, as shown in Fig.6 , in order not to violate the
observed gamma-ray background data by Fermi/LAT, which is in
agreement with the results in \citet{2013PhRvD..88l1301M}.

We also study how much the starburst galaxies  in different
luminosity ranges and redshift ranges contribute to the diffuse
gamma-ray background.  Fig.7 shows the contributions in each
luminosity range to the total flux. It can be seen that most flux
is contributed by the starbursts in the luminosity range from
10$^{11}$ to 10$^{13}$ $L_{\odot}$. For starburst galaxies of such
high luminosity, the pion production efficiency $f_\pi$ is close
to 1 and these galaxies are proton calorimeters.  Fig.8 shows the
contributions by starburst galaxies in different redshift ranges
to the total flux. As expected, the dominant contribution is by
starburst galaxies in the   redshift range of $1<z<2$.

\section{Discussions and Conclusions}
{Starburst galaxies, which have many supernova or hypernova
explosions},  are proposed to be a possible source for the
sub-PeV/PeV neutrinos recently detected by IceCube
\citep{2013PhRvD..88l1301M,2014PhRvD..89h3004L,2014arXiv1405.7648A}.
We have shown that the minimum diffuse gamma-ray flux associated
with these sub-PeV/PeV neutrinos is about $(2-3)\times10^{-8}{\rm
GeV cm^{-2} s^{-1} sr^{-1}}$, which is a factor of two lower than
the case without considering the synchrotron loss of the $e^\pm$
pairs  resulted from the absorption of very high energy photons.
This minimum diffuse gamma-ray flux constitute a fraction of
$\sim20\%$ of the observed gamma-ray background flux at $\sim100$
GeV energies, thus leaving a relatively large room for other
sources to contribute to the gamma-ray background.

It was proposed that hypernovae remnants in starburst galaxies
accelerate protons to $>10^{17}{\rm eV}$
\citep{2007PhRvD..76h3009W}, which then produce PeV neutrinos via
$pp$ collisions with the dense surrounding medium
\citep{2014PhRvD..89h3004L}. Remnants of normal supernovae may
accelerate protons to PeV energies, so they  may not contribute to
the $\ga100$ TeV neutrino flux,  but they can produce $<100$ TeV
gamma rays, thus contributing to the diffuse gamma-ray background
as well. But as long as the diffuse gamma-ray flux contributed by
these normal supernovae does not exceed that contributed by the
hypernova remnants too much, the total diffuse flux may still fall
below the observed background. It may be also possible that, if
the remnants of normal supernovae produce a gamma-ray flux 3-4
times higher, the total gamma-ray flux from starburst galaxies can
reach the level of the observed one. Interestingly,
\citet{Tamborra2014} recently show that star-forming and starburst
galaxies can explain the whole diffuse gamma-ray background in the
0.3-30 GeV range.

\acknowledgments We thank Peter M\'esz\'aros, Kohta Murase and
Ruoyu Liu for useful discussions, and the referee for the valuable
report. This work is supported by the 973 program under grant
2014CB845800, the NSFC under grants 11273016 and 11033002, and the
Excellent Youth Foundation of Jiangsu Province (BK2012011).

\newpage

\begin{figure*}[tbp]
 \epsscale{.9} \plotone{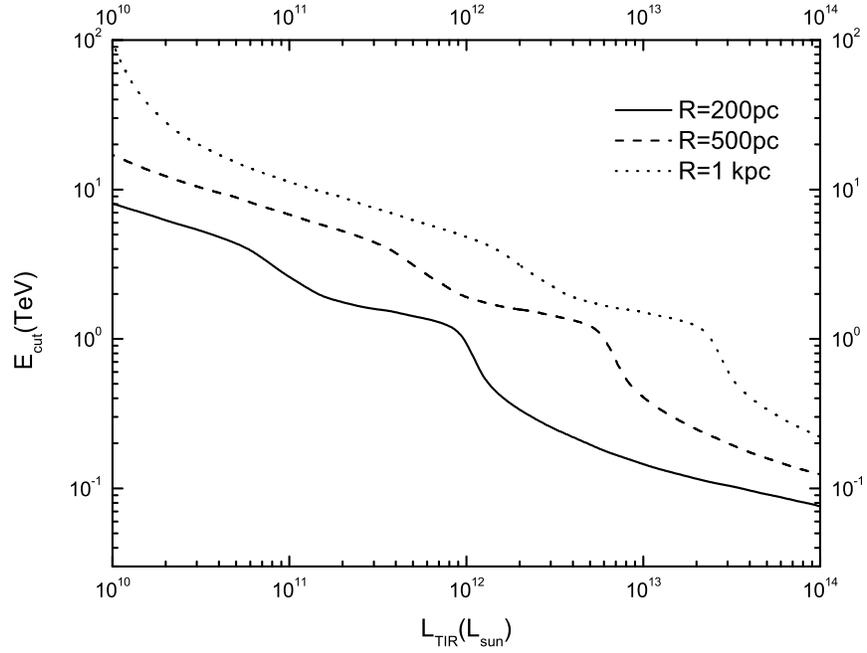}
\caption{The absorption cutoff energy of VHE gamma-rays by soft
photons in starburst galaxies with different $L_{\rm TIR}$. }
\label{ecut}
\end{figure*}

\begin{figure*}[tbp]
 \epsscale{.9} \plotone{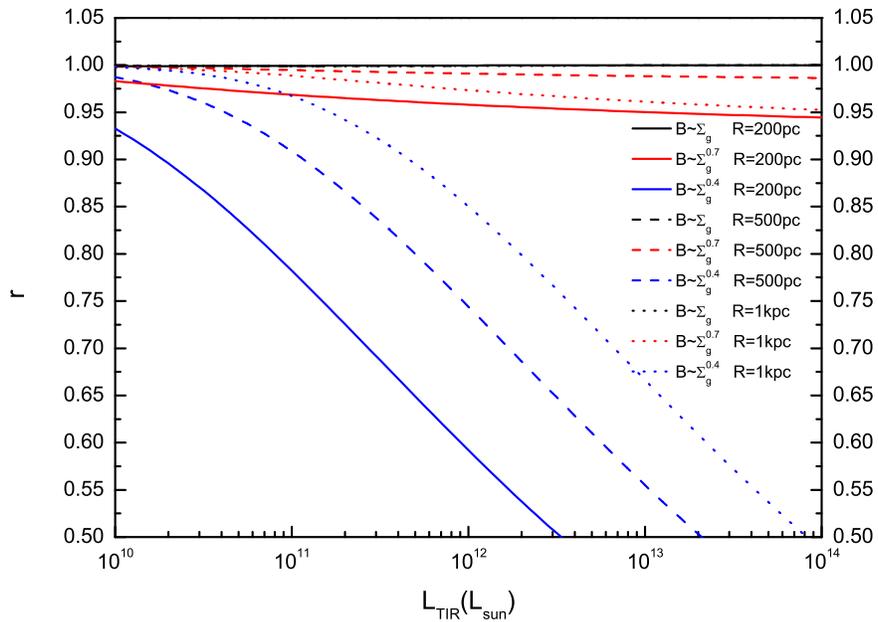}
\caption{The fraction of total energy loss of relativistic
$e^{\pm}$ through the synchrotron radiation  in the starburst
magnetic field. {The solid, dashed and dotted lines represent the
cases of $R=200$pc, $R=500$pc and $R=1$kpc respectively. The
black, red and blue lines show the cases of $B\propto \Sigma
_{g}$, $B\propto \Sigma _{g}^{0.7}$ and $B\propto \Sigma
_{g}^{0.4}$, respectively}. Note that the $B\propto \Sigma _{g}$
lines overlap with the horizontal line of r=1.} \label{ratio}
\end{figure*}

\begin{figure*}[tbp]
\epsscale{.9} \plotone{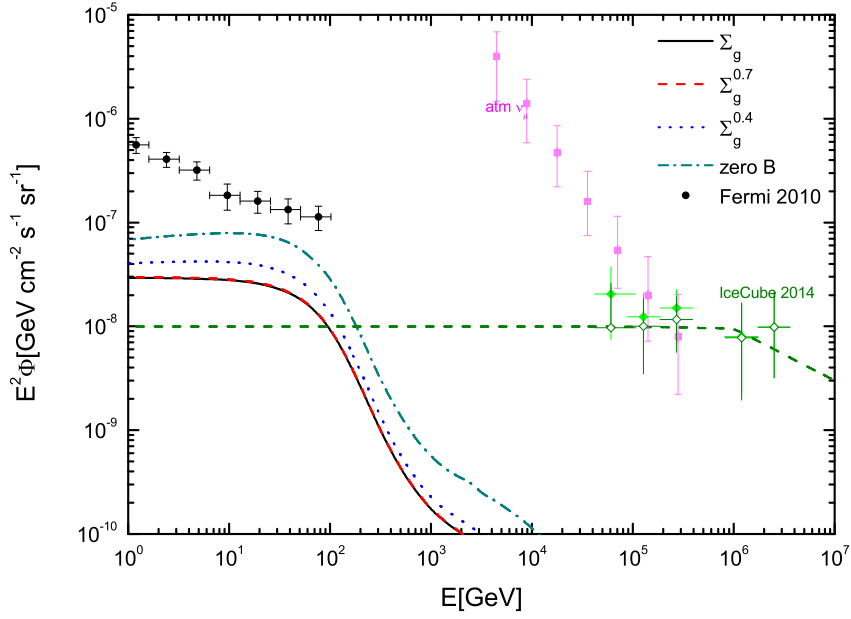} \caption{The {accumulated}
diffuse gamma-ray flux of starburst galaxies for different
assumptions of the magnetic fields in the starburst region.
$R=500{\rm pc}$ and $p=2$ are assumed. The black, red and blue
lines show the cases of $B\propto \Sigma _{g}$, $B\propto \Sigma
_{g}^{0.7}$ and $B\propto \Sigma _{g}^{0.4}$, respectively. For
illustration,  the green line shows the case of $B=0$. The
neutrino flux is obtained using Eq.(7). The extragalactic
gamma-ray background data from Fermi/LAT are depicted as the black
dots. The atmospheric neutrino data and the IceCube data are also
shown.} \label{total}
\end{figure*}

\begin{figure*}[tbp]
\epsscale{.9} \plotone{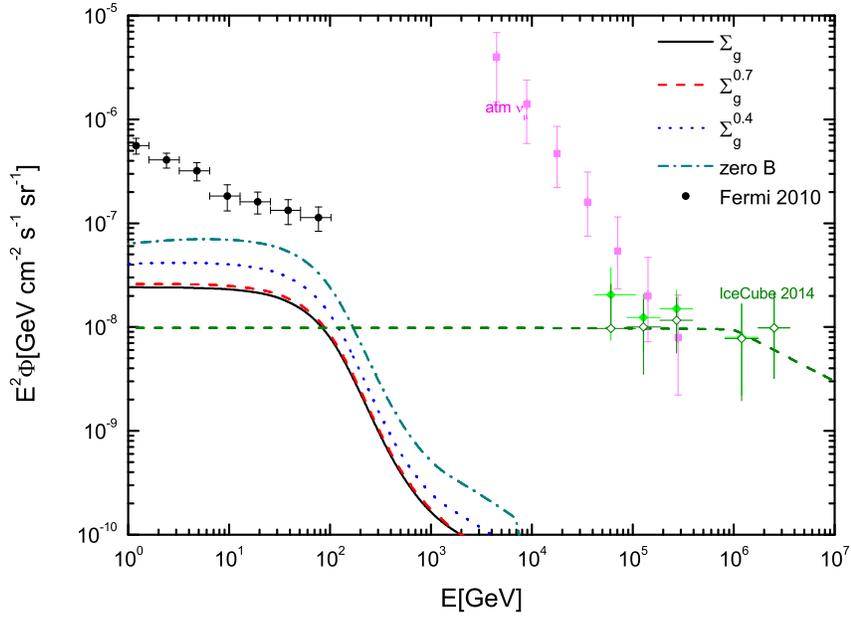} \caption{The same as {figure 3},
but with $R=200$pc.}
\end{figure*}

\begin{figure*}[tbp]
\epsscale{.9} \plotone{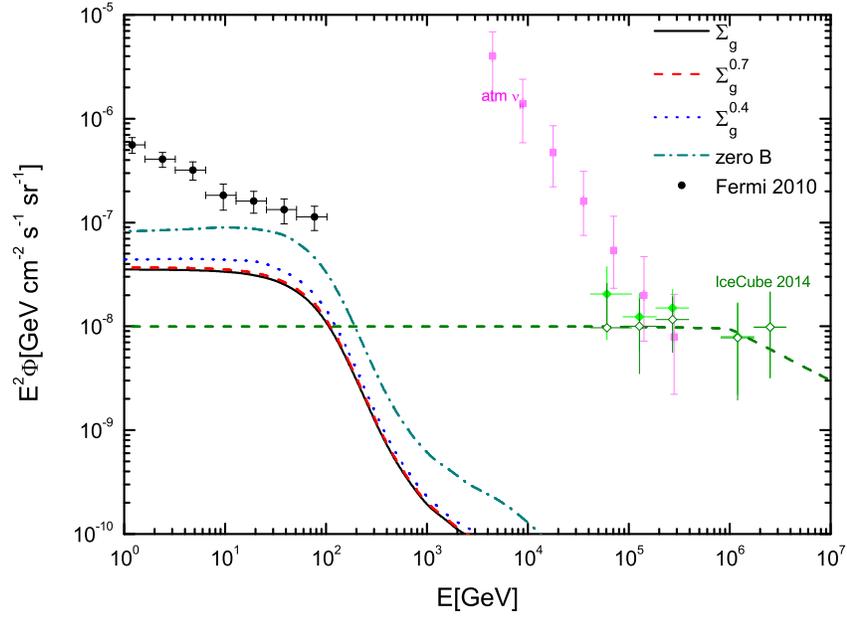} \caption{The same as {figure 3},
but with $R=1$kpc.}
\end{figure*}

\begin{figure*}[tbp]
\epsscale{.9} \plotone{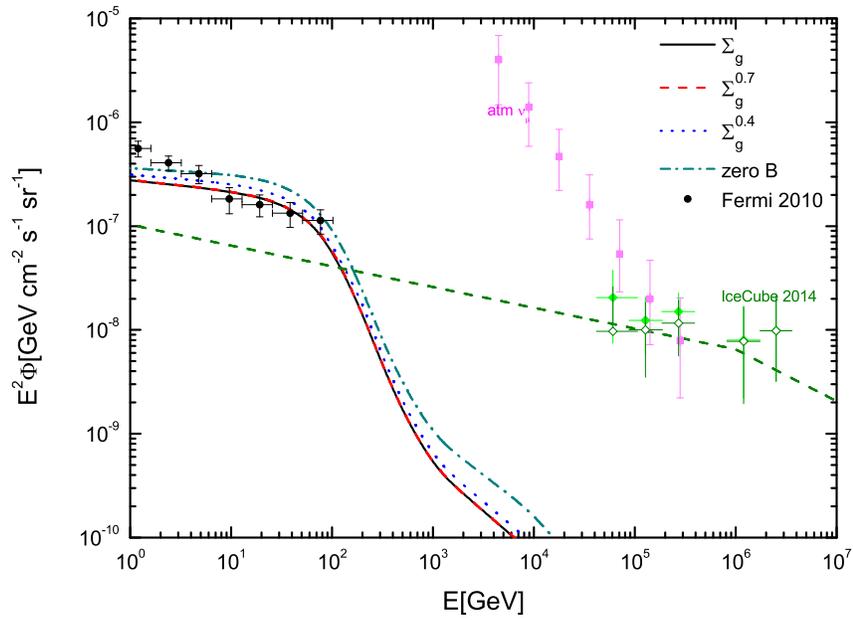} \caption{The same as {figure 3},
but assuming a steeper proton spectrum with $p=2.18$}
\end{figure*}

\begin{figure*}[tbp]
\epsscale{.9} \plotone{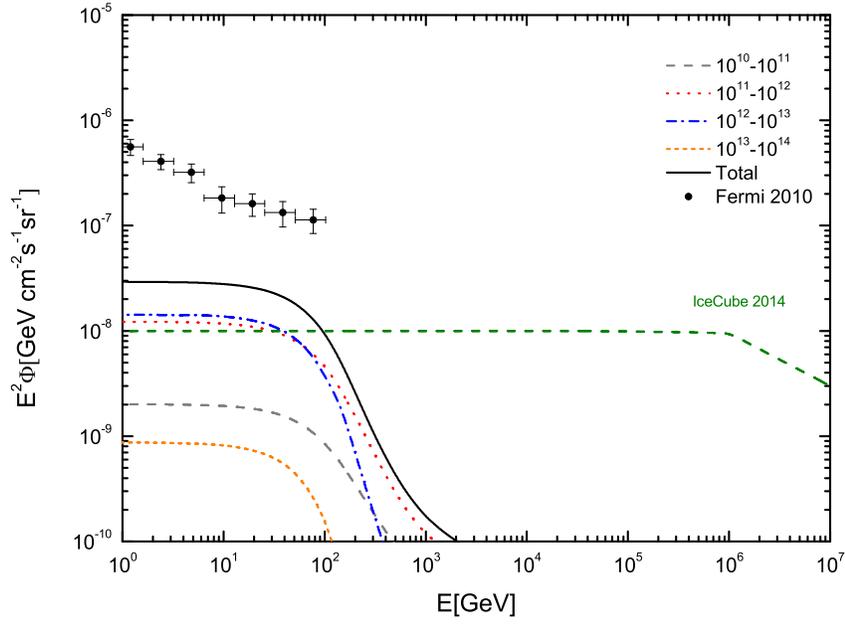} \caption{The diffuse gamma-ray
flux contributed by starburst galaxies   in different luminosity
ranges. The grey, red, blue and orange lines represent the
contributions by the starburst galaxies in the luminosity ranges
of $10 ^{10}-10^{11}L_{\odot }$, $10^{11}-10^{12}L_{\odot }$,
$10^{12}-10^{13}L_{\odot }$, and $10^{13}-10^{14}L_{\odot }$,
respectively. The black line represents the sum of them.
$R=500{\rm pc}$, $B\propto \Sigma _{g}$ and $p=2$ are assumed.}
\label{lbin}
\end{figure*}

\begin{figure*}[tbp]
\epsscale{.9} \plotone{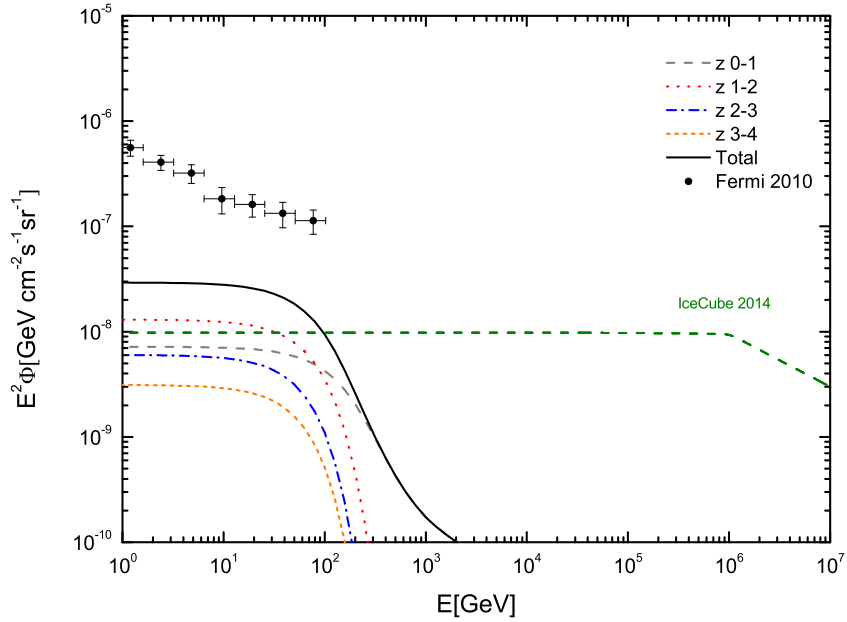} \caption{The diffuse gamma-ray
flux contributed by starburst galaxies   in different redshift
ranges. {The grey, red, blue and orange} lines represent the
contributions   by the starburst galaxies in the redshift ranges
of $0-1$, $1-2$, $2-3$ and $3-4$, respectively. The black line
represents the sum of them. $R=500{\rm pc}$, $B\propto \Sigma
_{g}$ and $p=2$ are assumed.} \label{zbin}
\end{figure*}

\end{document}